# Anion and cation emission from water molecules after collisions with 6.6-keV $^{16}O^+$ ions

Z. Juhász,[1] B. Sulik,[1] E. Lattouf,[2] E. Bene,[1] B. A. Huber,[2] P. Herczku,[1] S. T. S. Kovács,[1] A. Méry,[2] J.-C. Poully,[2] J. Rangama,[2] J. A. Tanis,[2,3] V. Vizcaino,[2] and J.-Y. Chesnel[2,*]

[1]*Institute for Nuclear Research, Hungarian Academy of Sciences (MTA Atomki), H-4001 Debrecen, P.O. Box 51, Hungary*

[2]*Centre de Recherche sur les Ions, les Matériaux et la Photonique, Normandie Université, ENSICAEN, UNICAEN, CEA, CNRS, CIMAP, 14000 Caen, France*

[3]*Department of Physics, Western Michigan University, Kalamazoo, Michigan 49008, USA*

*Corresponding author: jean-yves.chesnel@ensicaen.fr

Anion and cation emission following water dissociation was studied for 6.6-keV $^{16}O^+$ + $H_2O$ collisions. Absolute cross sections for the emission of all positively and negatively charged fragments, differential in both energy and observation angle, were measured. The fragments formed in hard, binary collisions appearing in peaks were distinguishable from those created in soft collisions with many-body dynamics that result in a broad energy spectrum. A striking feature is that anions and cations are emitted with similar energy and angular distributions, with a nearly constant ratio of about 1:100 for $H^-$ to $H^+$. Model calculations were performed at different levels of complexity. Four-body scattering simulations reproduce the measured fragment distributions if adequate kinetic-energy release of the target is taken into account. Providing even further insight into the underlying processes, predictions of a thermodynamic model indicate that transfer ionization at small impact parameters is the dominant mechanism for $H^+$ creation. The present findings confirm our earlier observation that in molecular fragmentation induced by slow, singly charged ions, the charge states of the emitted hydrogen fragments follow a simple statistical distribution independent of the way they are formed.

## I. INTRODUCTION

The interaction between ions and molecular targets is an essential process in several areas of physics, biology, and chemistry. Understanding ion-induced molecular fragmentation and its dynamics is a challenging problem, both experimentally and theoretically. The dissociation pathways and the energy and angular distributions of fragments ejected from the target molecule depend strongly on the energy and momentum transferred by the projectile to each target electron and nucleus, and on whether the collision involves excitation or electron removal via direct ionization, electron capture, or transfer ionization.

Collision-induced fragmentation of water molecules by ion impact has received considerable attention in the last decades [1-16] mainly because it is not only of fundamental interest but also important for many applications. Water is the main constituent of biological tissues and it is prominently present in the atmosphere of comets, moons, and planets. The wide range of circumstances in which ions interact with water molecules makes the study of collision-induced fragmentation of water essential in various fields of applied science, such as astrophysics, atmospheric research, biophysics, and plasma physics.

The pioneering works of Werner *et al.* [2,3] investigated the fragmentation of isolated water molecules by the impact of bare and hydrogen-like ions with incident energies ranging from a few tens to a few hundreds of keV/nucleon. These works provided total and partial fragmentation cross sections as well as partial cross sections for the production of individual cationic fragments. The total kinetic energy distribution of the ionized fragments was found to depend strongly on projectile species and incident energy. Gobet *et al*. [4,5] and Luna *et al.* [6,7] extended these studies by providing partial cross sections for fragmentation following electron capture and direct ionization events. More recent studies [17-24] gave evidence for multiple electron processes in $H_2O$ fragmentation by proton and $He^+$ impact at energies ranging from 20 keV/nucleon to several MeV/nucleon. Other groups [10-16] focused investigations on the regime of lower incident energies, down to keV/nucleon and sub-keV/nucleon ranges. In this regime, measurements of absolute cross sections for fragment emission, differential in both emission energy and angle, provided further insight into the fragmentation mechanisms of Coulomb explosion and binary collisions [11-13,25-29].

The charged fragments investigated in the aforementioned studies are cations. For other molecular targets, a few studies reported ion-induced hydrogen anion formation in gentle, low-momentum transfer collisions [30-35]. Recently, we have shown that anion creation is a more general process in molecular fragmentation induced by cation impact [36-38]. We found that anions are also created in hard collisions involving energetic encounters between two atomic cores (a "binary-encounter process") at few-keV impact energies. Hydrogen anions were observed in the fragmentation of both H-containing projectile ions (in 7-keV $OH^+$ + Ar and $OH^+$ + acetone collisions [36,37]) and target molecules (in 7-keV $OH^+$ + acetone [36] and 6.6-keV $O^+$ + $H_2O$ [38] collisions). In our work on $H_2O$ fragmentation [38], full energy and angular distributions of the emitted anions were measured and analyzed in terms of a simple classical four-body Monte Carlo model. The relative contributions of the binary and many-body processes to anion production were also determined.

The emission of energetic $H^-$ fragments has shown a simple statistical character in all the above cases. For collisions involving $OH^+$ projectiles, the angular-dependent yield of the $H^-$ fragments from the projectile is proportional to the calculated angle-differential cross section for the elastic scattering of an H atom on the target atom(s). In $OH^+$ + acetone and in $O^+$ + $H_2O$ collisions, the yields of fast $H^-$ fragments from the target are proportional to the calculated recoil yields in $O^+$ + H elastic scattering. A common feature of the production of fast hydrogen fragments in binary-encounter events was that the angle-differential cross sections of $H^-$ anions and $H^+$ cations showed very similar angular dependence, as they are practically proportional to each other. This result suggests that the charge states of the outgoing high-velocity H fragments are statistically populated.

A primary goal of this paper is to address the question of whether this statistical aspect holds for many-body processes, too. Here, we complement our previous study [38] devoted to the formation of anionic fragments from water molecules colliding with 6.6-keV $O^+$ ions. We perform a comparative study of both anion and cation production following collision-induced molecular fragmentation. Specifically, our earlier comparison between single-differential cross sections (SDCS) for $H^+$ and $H^-$ formation in binary-encounter events [37,38] is extended to the comparison of *double-differential* cross sections (DDCS) for $H^+$ and $H^-$ formation in both cases of binary and many-body processes. We show that both the kinetic-energy and angular distributions of the $H^+$ fragments are very similar to those of the ejected $H^-$ ions, independently of whether these fragments are formed via binary or many-body processes.

Another goal of this paper is to gain further insight into the processes behind the observed fragmentation spectra by using a recently developed statistical-type thermodynamic model [39] to calculate absolute DDCS for fragment emission and to compare with the experimental results. We adapted the model for the treatment of the triatomic $H_2O$ target taking into account its initial vibrational motion. The good overall agreement between model calculations and experiment over the entire angular and energy range allows for separate identification of different fragment-ion formation mechanisms and of the associated electronic processes.

## II. EXPERIMENTAL METHOD

The molecular fragmentation experiments were performed with a conventional crossed-beam setup installed at the ARIBE beamline at the Grand Accélérateur National d'Ions Lourds (GANIL) in Caen, France. The beam of 6.6-keV $^{16}O^+$ was supplied by an electron cyclotron resonance (ECR) ion source with a typical beam current of ~150 nA measured in a two-stage Faraday cup. Following guiding into the experimental chamber, the beam passed through a jet of water vapor with a target density estimated to be about $10^{11}$ molecules/cm$^3$. Magnetic shielding reduced the Earth's magnetic field to a few milligauss inside the chamber.

Both negatively and positively charged fragments originating from the collision region were analyzed with kinetic energy by means of a single-stage, 45° electrostatic parallel-plate spectrometer (see Fig. 1). The spectrometer was mounted on a rotatable ring, thus allowing measurements at different observation angles ($\theta$) relative to the incident ion beam direction, with an angular acceptance of 2°. The energy resolution of the spectrometer was 5%. Further details of the experimental method and data analysis can be found in Refs. [37,40,41].

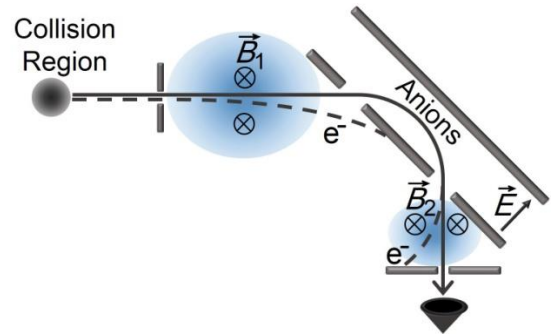

**Figure 1.** (Color online) Scheme of the electrostatic analyzer equipped with magnetic filters at its entrance and exit. Electrons are deflected so that only anions reach the detector. (Reprinted from Ref. [38] with permission).

In order to avoid the undesired electron contributions in the anion fragment spectra, magnetic filters were applied at two points along the path of the fragments traveling through the spectrometer (see Fig. 1). Electrons emerging from the collision region were deflected by applying a $10^{-4}$ T magnetic field $\vec{B}_1$ in front of the entrance slit of the analyzer. This field prevented electrons up to 2 keV from entering the analyzer. Secondary electrons originating from the analyzer were filtered out by an additional magnetic deflection field $\vec{B}_2$ between the exit slit of the analyzer and the Channeltron detector. Each of the magnetic filters was provided by a pair of coils, designed to minimize the stray magnetic fields inside the analyzer. Doubly differential fragment-ion emission cross sections both for cations and anions have been determined by a standard normalization procedure given in Ref. [37].

In order to obtain the relative contributions of the different fragment-ion species, complementary time-of-flight (TOF) measurements were performed using the same apparatus. These TOF measurements were also used to confirm the sufficient reduction of electron contamination in the low-energy anion spectra. The incident ion beam was pulsed with a period of 60 µs and pulse duration of 2 µs. The time of flight of the charged fragments from the collision region to the Channeltron detector was measured by setting the spectrometer to a selected detection energy. The flight distance was ~200 mm. At this distance, the pulse duration of 2 µs was short enough to separate the hydrogen ions from the oxygen-containing ions at emission energies ranging from 3 to 25 eV. To keep a satisfactory signal-to-noise ratio, no attempt was made to reduce the pulse duration below 1 µs for a separate identification of the O, OH, and $H_2O$ components of the detected oxygen-containing ions.

## III. RESULTS

### A. DDCS for anion and cation emission

Figures 2(a) and 2(b) show experimental double-differential cross sections for anion and cation emission, respectively, in 6.6-keV $^{16}O^+ + H_2O$ collisions. These differential cross sections are displayed as a function of the fragment kinetic energy (per charge unit) for various observation angles with respect to the beam direction. Fragments emitted in the forward (resp., backward) direction are observed at angles lower (resp., larger) than 90°.

The main component of each spectrum is a broad, slowly decreasing structure, which we refer to as the *continuous spectrum*. For both anion and cation emission, the continuous spectra maximize at low emission energies. There, the relative contributions of the light H ions and the heavy oxygen-containing fragments (ionic O and OH species) were derived from TOF measurements. In Figs. 2(a) and 2(b) these H and O contributions are depicted by means of red circles and green diamonds, respectively. Comparison of these data reveal the following features:

(i) For both anion and cation emission, the relative contribution due to H ions is larger than that due to O-containing fragments.

(ii) The low-energy $H^+$ and $H^-$ fragments are emitted with very similar energy and angular dependences. This is also the case for positively and negatively charged oxygen-containing fragments.

(iii) The angular distribution of slow oxygen fragments (both anions and cations) is anisotropic, with a maximum centered near 85°–90°. This behavior has been reported for binary collisions, where the slowest recoil ions move almost perpendicularly to the projectile trajectory. This is consistent with the picture that the collision of the projectile with the target oxygen atom is close to a pure binary encounter, in which the light hydrogen atoms play a minor role.

(iv) The angular distribution of slow hydrogen fragments (both $H^+$ and $H^-$) is nearly isotropic. This finding indicates that emission of these light fragments is far from being exclusively governed by binary collisions with the heavy projectile.

We note that the role of many-body dynamics in low-energy H emission is expected to be significant due to the perturbation induced by the heavy target oxygen atom on the trajectories of light H fragments. This picture is supported by the fact that the $H^+$ and $H^-$ energy spectra resemble those of slow electrons emitted in ion-atom collisions [42].

At higher emission energies, pronounced peaks appear in both anion and cation spectra [Figs 2(a) and 2(b)], but only at forward observation angles (< 90°). The mean energy of these peaks decreases strongly with increasing angle $\theta$, following a $\cos^2(\theta)$ law. From a kinematic calculation assuming an elastic two-body collision between the projectile and a single target atom, the peak at lower energy (in red) is assigned to $H^-$ [Fig. 2(a)] or $H^+$ [Fig. 2(b)] and the other peaks (in green) to oxygen-containing anions [Fig. 2(a)] or cations [Fig. 2(b)]. These ions are recoil ions formed in hard binary collisions occurring at small impact parameters with a large momentum transfer. As one of the receding target atoms leaves the collision complex it may either capture enough electrons to become an anion (or to be neutral) or lose electrons to become a cation.

Since the kinetic energy of recoil fragments formed in hard two-body collisions barely depends on their final charge state, the measurement of their energy per charge unit allows for identification of singly and doubly charged oxygen cations, as shown in Fig. 2(b). We note that in a binary O-O collision the *scattered projectile* may also either capture electrons or be ionized, and thus contribute to the oxygen peaks, both in the anion and cation spectra.

The mass distribution of the cationic oxygen-containing species (O, OH, $H_2O$) may differ from that of the anionic ones. However, as noted above, the present experiment was not dedicated to the measurement of the mass distribution of the heavier oxygen species. Thus, in this paper, no attempt is made for a further detailed comparison between the cationic and anionic oxygen contributions.

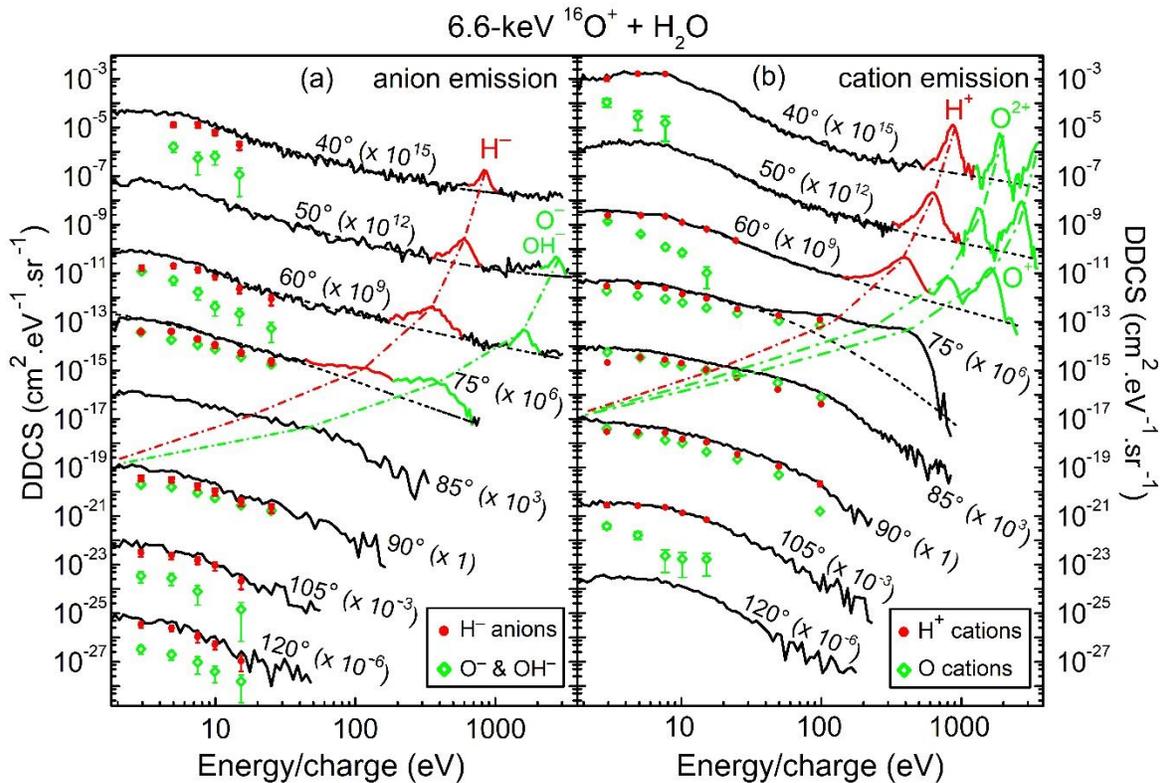

**Figure 2.** (Color online) Full curves: DDCS for anion (a) and cation (b) emission in 6.6-keV $^{16}O^+$ + $H_2O$ collisions at the indicated observation angles. These cross sections are doubly differential in emission energy and solid angle. For graphical reasons, each spectrum is multiplied by the indicated factor. Fragments formed in many-body collisions are observed in the slowly decreasing "continuous" part of the spectra (full and dashed curves in black). Fragments created in binary-encounter collisions appear in peaks. The shifts of the peak centroids toward lower energies at larger angles are graphically emphasized by the colored dash-dotted curves. The points at lower energies give the relative contributions due to hydrogen ions (red circles) and oxygen-containing ions (green diamonds) from TOF measurements.

## B. Comparison of $H^+$ and $H^-$ energy distributions

The similarity between the energy distributions of $H^+$ and $H^-$ fragments is further highlighted in Fig. 3. This figure shows the DDCS for anion and cation emission in 6.6-keV $^{16}O^+$ + $H_2O$ collisions at observation angles of 60° (top) and 75° (bottom).

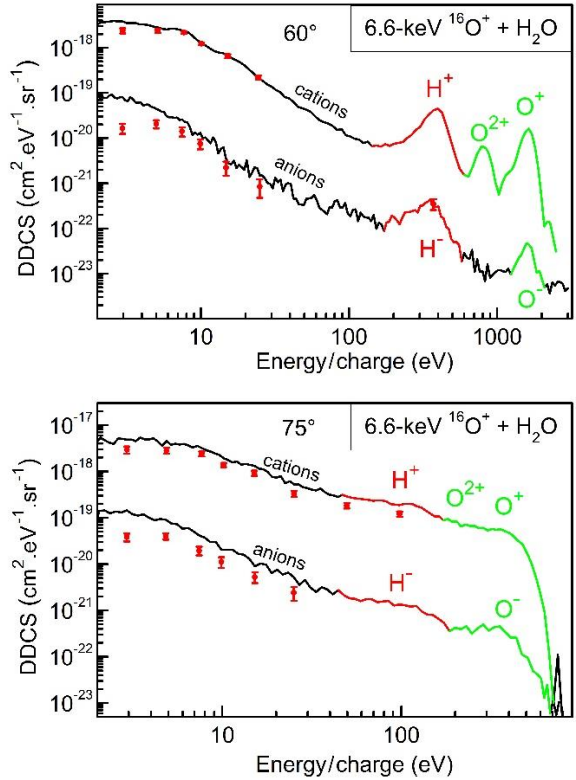

**Figure 3.** (Color online) Full curves: DDCS for anion and cation emission in 6.6-keV $^{16}O^+$ + $H_2O$ collisions at the observation angles of 60° (top) and 75° (bottom). The points (red circles) at lower energies show the relative contributions due to hydrogen fragments ($H^+$ or $H^-$).

At each observation angle, the energy distributions of cations and anions are largely similar as the DDCS for $H^+$ and $H^-$ production are about proportional. The $H^-/H^+$ ratio is in the range of about 0.5%–1% over the entire emission energy range investigated here. As can be seen in Fig. 3 when comparing data at 60° and 75°, the energy distributions at different angles can differ significantly. The close similarity of the $H^-$ and $H^+$ spectra, however, holds at all the angles explored here (Figs. 2 and 3). In the next subsection, we illustrate quantitatively this feature by comparing the $H^+$ and $H^-$ angular distributions at selected emission energies.

## C. Comparison of $H^+$ and $H^-$ angular distributions

In our previous studies on different molecular collisions at the same velocity [37,38], we pointed out the similarity between $H^+$ and $H^-$ emission in the particular case of hydrogen emission in hard binary-encounter collisions involving large momentum transfer to H cores at small impact parameters. In the present study, we extend the comparison between $H^+$ and $H^-$ emission to the case of soft many-body collisions involving low-momentum transfer at large impact parameters.

For these soft collisions, the top part of Fig. 4 shows the angular distribution of the DDCS for $H^+$ and $H^-$ formation at the three emission energies of 5, 7.5, and 10 eV. The bottom part of Fig. 4 displays the ratio between the $H^-$ and $H^+$ DDCS shown in the top part of the figure. At each emission energy both $H^-$ and $H^+$ angular distributions are only weakly anisotropic with a broad maximum around 80°–90°. At energies lower than ~10 eV, the $H^-$ and $H^+$ DDCS vary by less than a factor of 3 in the 40°–140° angular range.

The similarity between the $H^-$ and $H^+$ angular distributions is further highlighted by the fact that the $H^-/H^+$ DDCS ratio is constant within the error bars at a given emission energy, equal to about 0.5%–1%. The same conclusion applies at larger emission energies, such as 15 and 25 eV (not shown), but lower cross sections at these energies led to lower count rates, and thus, to larger error bars. These results show that the $H^-/H^+$ DDCS ratio barely depends on the emission energy and angle in the case of soft many-body collisions.

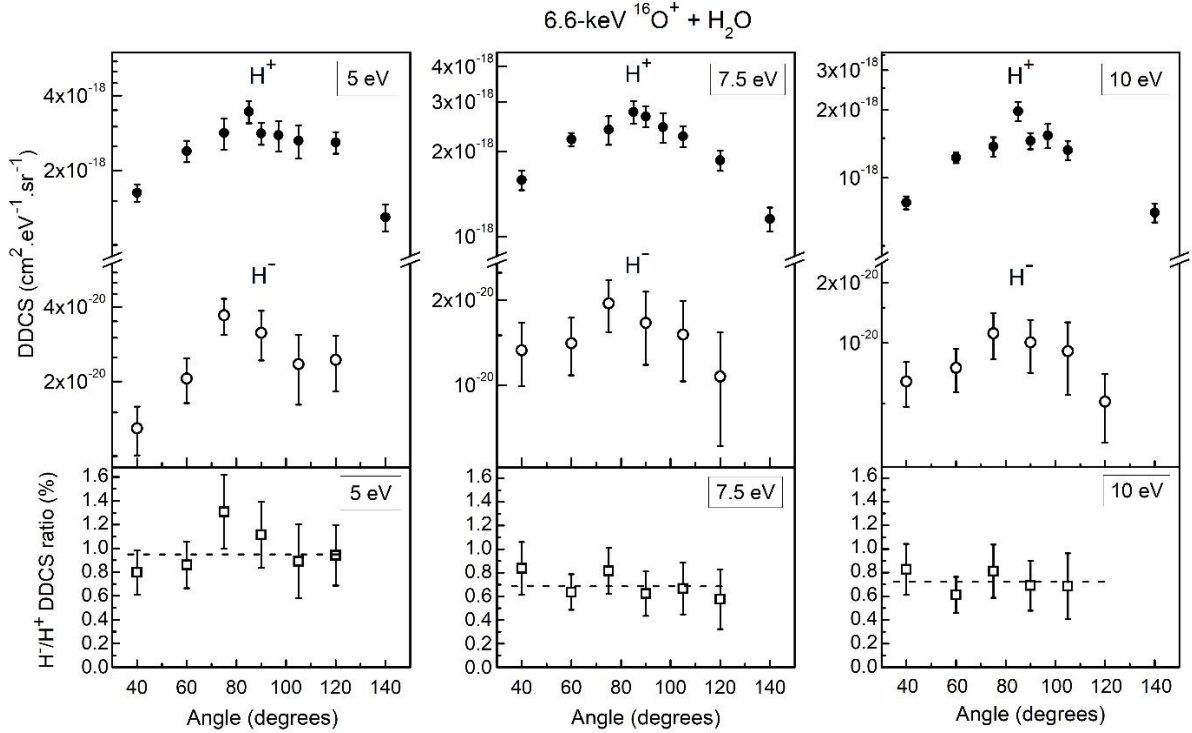

**Figure 4. Top:** Angular distribution of the DDCS for H$^−$ (open circles) and H$^+$ (full circles) emission in 6.6-keV $^{16}$O$^+$ + H$_2$O collisions at the energies of 5 eV (left), 7.5 eV (middle) and 10 eV (right). Only *relative* error bars due to statistical uncertainties are shown. **Bottom:** Ratios between the experimental H$^−$ and H$^+$ DDCS shown in the top part.

On the other hand, hard binary-encounter collisions lead to the emission of fast H$^+$ and H$^−$ recoil ions (Figs. 2 and 3). In Fig. 5, we recall our previously published results on the angular dependence of single-differential cross sections (SDCS) for H$^−$ and H$^+$ emission from H$_2$O via binary processes [38]. The angular dependence of these cross sections is compared with the theoretical cross section (see Ref. [36]) for the elastic recoil of an H atom by the impact of an O atom [Fig. 5(a)]. The theoretical curves match the experimental data when multiplied by appropriate factors. These factors represent the relative populations of the different charge states of H fragments. It was found that $(0.7 \pm 0.4)\%$ of the ejected H atoms becomes H$^−$ and $(60 \pm 33)\%$ of them becomes H$^+$. The SDCS H$^−$/H$^+$ ratio is about constant, with an average value of ~1% [Fig. 5(b)].

The present results for H$^−$ emission via many-body processes (Fig. 4) generalize our recent findings for H$^−$ emission via the binary process in 6.6-keV O$^+$ + H$_2$O collisions [38], as well as in other molecular collisions at the same velocity [36,37]. Namely, the relative populations of the different charge states of the hydrogen fragments do not depend significantly on the emission angle, the impact parameter, or the momentum transferred between the collision partners. This finding suggests that the charge-state distribution of the hydrogen fragments is akin to a statistical distribution, whether these fragments are formed via the binary process or via many-body processes. This statistical aspect likely stems from the fact that the number of possible final charge states for the outgoing fragment is very limited compared to the number of electronic transitions which may occur in each collision.

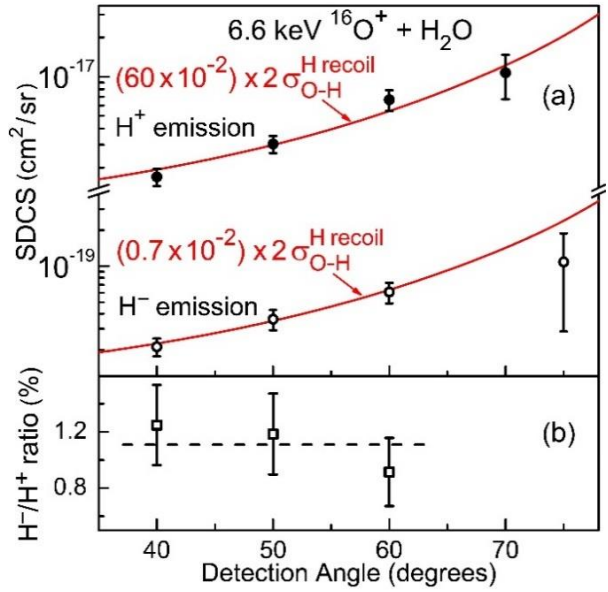

**Figure 5.** (Color online) **(a)** SDCS for H⁻ (open circles) and H⁺ (full circles) emission via the binary-encounter process as a function of the observation angle. These cross sections are differential in the emission solid angle. Only *relative* error bars due to statistical uncertainties are shown, except the rightmost points, where a larger uncertainty stems from the overlap of peak structures. Red curves: calculated SDCS for two-body elastic recoil of H atoms by 6.6-keV oxygen impact, multiplied by factors representing the fraction of the different charge-state components. **(b)** Ratio between the experimental H⁻ and H⁺ SDCS reported in (a). (Reprinted from Ref. [38] with permission).

### D. Total cross sections

Total cross sections are obtained by integrating the double-differential cross sections over energy and solid angle (Fig. 2). The total fragmentation cross sections for anion and cation formation in 6.6-keV $^{16}O^+ + H_2O$ collisions are found to be $(5 \pm 3) \times 10^{-18}$ cm$^2$ and $(6 \pm 3) \times 10^{-16}$ cm$^2$, respectively. In principle, this evaluation requires an accurate measurement of the contribution due to the slowest fragments, but this is challenging as their detection may be altered by stray fields. Here, we extrapolated the energy spectra by assuming a constant DDCS in the energy range from 0 to 2 eV. In this case, the relative contribution of the extrapolated part is about 10 %. This value may be considered as a lower limit since formation of sub-eV recoil fragments in soft binary collisions may result in higher DDCS values near 90°. After integration over energy, the cross section, differential in solid angle, is found to be anisotropic, with a maximum at about 90°. This is true for both anion and cation emission. If isotropic, fragment emission in the angular range from 40° to 140° would represent 77% of the total cross section. But here, the observed anisotropy makes the relative contribution of this angular range larger, on the order of 90%.

For both anion and cation emission, the TOF measurements show that about two-thirds of the detected fragments are hydrogen ions. Hence, the total cross section for producing H⁻ ions in the present collision is estimated to be ~$3 \times 10^{-18}$ cm$^2$, while it is ~$4 \times 10^{-16}$ cm$^2$ for H⁺ ions. By integrating the continuous part of the spectra over energy and angle, we find that ionic fragment formation via many-body processes accounts for about 90% of the total cross section, independently of whether the ejected fragments are positively or negatively charged.

Total cross sections for dissociative electron attachment (DEA) to water in the gas phase were previously measured as a function of the kinetic energy of the incident electrons [43]. The present data show that cross sections for cation-induced anion emission from H$_2$O can be as large as those for anion formation by slow electron impact at incident energies of a few eV.

## IV. MODEL RESULTS AND DISCUSSION

The multielectronic character of collision processes impedes the development of a rigorous theory of molecular fragmentation in ion-molecule collisions. We thus performed numerical simulations to interpret the experimental data. In a first approach, we derived fragment emission yields from simplistic four-body scattering simulations. The method is the same as the one used in Ref. [38]. Then, in a second approach, we applied the statistical-type thermodynamic model developed in Ref. [39] to calculate absolute DDCS for fragment emission and to compare with the experimental results. Here, since the oxygen-containing species (O, OH, $H_2O$) are not separately identified experimentally, the comparison between simulations and experiment focuses only on the emission of $H^+$ and $H^-$ fragments.

### A. Four-body scattering simulations

We performed numerical simulations of the trajectories of the fragments by assuming a two-body interaction between each pair of atoms. For each pair, an *ab initio* calculation using the MOLPRO code [36,44] provided the interaction potential as a function of the internuclear distance. The two-body potentials refer to the ground-state energy of the diatomic subsystems as a function of the interatomic distance. The trajectories of the four atomic cores of the present collision system were calculated by solving the coupled Newton equations of motion. The initial position of the projectile and the initial orientation of the target molecule were randomly defined. Moreover, each target atom was assumed to be initially at rest and the initial O-H distance and H-O-H angle (in $H_2O$) were set equal to their equilibrium values. Hence, the model has four random parameters to define the initial conditions of each simulated collision. The energy and angular distributions of the different atomic cores were then determined by simulating a large number of collisions ($\sim 3 \times 10^6$).

In principle, these numerical simulations can only provide insights into the dynamics of the atomic cores, as they do not provide information about their final charge state. However, we took advantage of the fact that the anion and cation signals are about proportional to extract from these simulations approximate cross sections for emission in a particular charge state. To do so, we multiplied the cross sections obtained from the simulation by an appropriate constant factor that depends on the charge state of the emitted atomic cores. To ensure that the intensity of the calculated binary peaks matches the experimental results, we multiplied the calculated hydrogen emission cross sections by either the $H^-$ fraction ($0.7 \times 10^{-2}$) or the $H^+$ fraction ($60 \times 10^{-2}$) derived from Fig. 5.

For an initial test, we performed a simulation in which no electronic excitation was taken into account. The $H^-$ and $H^+$ emission DDCS obtained in this test calculation are displayed with dotted curves in Figs. 6(a) and 6(b), respectively. First, for both $H^-$ and $H^+$ emission, the calculated high-energy peak is narrower than the experimental one. This disagreement suggests that excitation and ionization processes cannot be neglected as they may affect the energy (and angular) distribution of both $H^-$ and $H^+$ fragments formed in hard binary-encounter collisions. Secondly, at lower emission energies, both calculated $H^-$ and $H^+$ cross sections are significantly smaller than the experimental ones. Indeed, in the simulated soft collisions occurring at impact parameters larger than $\sim 1$ atomic unit (a.u.) (relative to the active H core), the kinetic energy transferred to the H cores is generally lower than the dissociation energy of the OH bonds of $H_2O$ in its ground state. Consequently, this test simulation led to the result that less than 10% of the OH bonds are broken in such soft collisions, thus leading to underestimated cross sections at the lowest emission energies. This result suggests that the formation of dissociative states is required for an efficient OH bond breaking in soft collisions at large impact parameters.

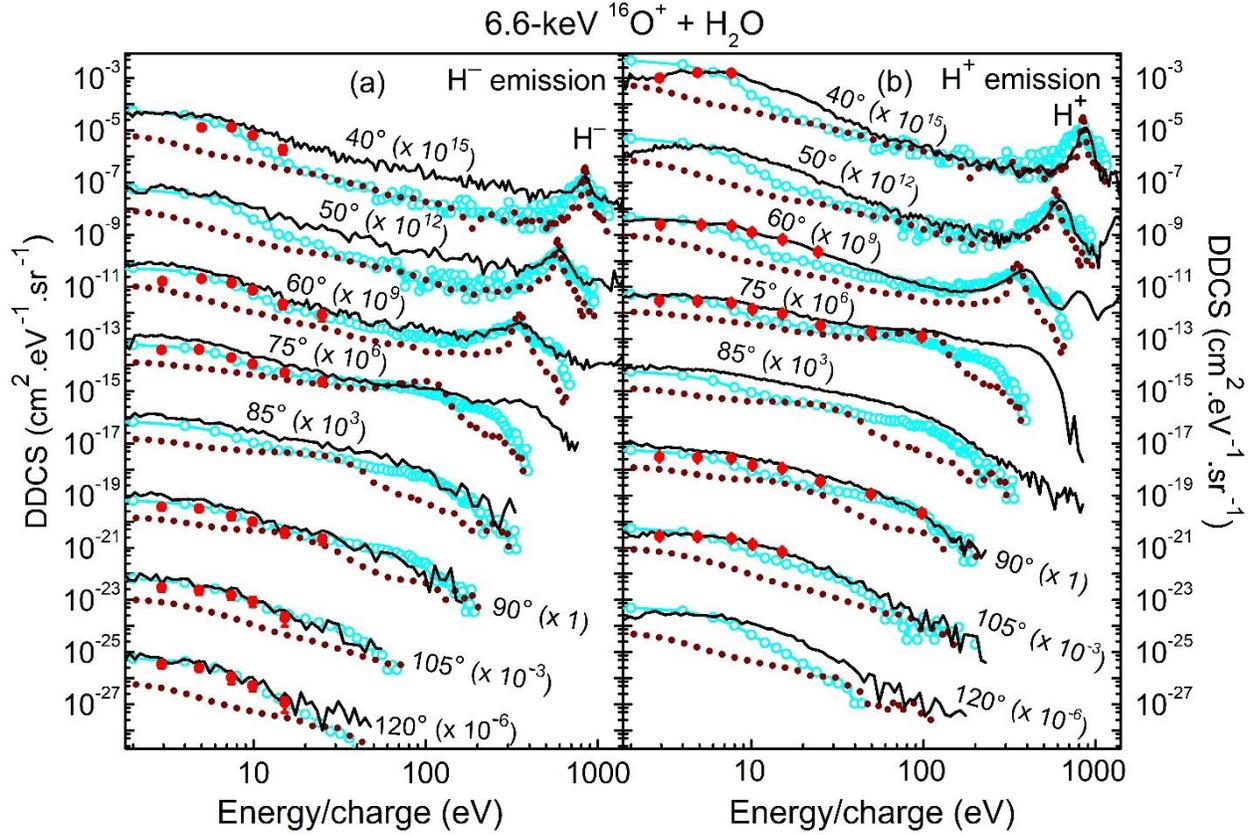

**Figure 6.** (Color online) Comparison between experimental DDCS and four-body scattering simulations for hydrogen fragment emission at the indicated observation angles in 6.6-keV $^{16}O^+$ + $H_2O$ collisions. **Open circles:** Simulated energy distribution of the $H^-$ (a) and $H^+$ (b) fragments at different angles with KER. **Dotted curve:** Same without KER. **Full curve:** Experimental DDCS for anion (a) and cation (b) emission. **Red, full circles:** Experimental $H^-$ (a) and $H^+$ (b) contributions from TOF measurements. Each spectrum is multiplied by the indicated factor.

Electronic processes such as dissociative excitation and ionization may result in kinetic-energy release to the atomic cores of the molecular target [45]. Therefore, we performed an additional and more realistic simulation by introducing kinetic-energy release (KER) into the model for the closest-approach collisions with impact parameter ranging from 0 to 3 a.u. Since a large variety of excitation processes can occur in energetic collisions involving many atoms, the number of dissociative channels can be extremely large. Hence, the KER is likely to be of statistical character. Previous measurements on ion-induced dissociation of $H_2O$ molecules show that the kinetic energy released in the cleavage of an OH bond is on the order of 5 eV and can even exceed 20 eV when highly excited states are involved [26]. To reasonably match these previous data in our simulations, we made the assumption that the KER ($\approx \frac{1}{2} m_H v_{KER}^2$) is a random variable with a Gaussian distribution centered at 5 eV with a standard deviation of 4 eV, independently of the impact parameter $b$. Hence, the model has three adjusting parameters, *i.e.*, the condition $b \leq 3$ a.u. for introduction of KER and the mean and the standard deviation of the KER distribution. As negative KER values were excluded here, the KER value ranged from 0 to 10 eV in 88% of the simulated collisions, while KER values larger than 15 eV occurred in less than 0.7% of the simulated events. As in [37,38], KER is introduced by adding a velocity component along the OH axis, $\vec{v}_{KER}$, to the velocity of the H atom when the distance between the projectile and the active H atom is minimum.

The $H^-$ and $H^+$ emission DDCS obtained in this latter simulation are displayed by means of open circles in Figs. 6(a) and 6(b), respectively. Though only a qualitative description was expected from this simulation, it appears that the introduction of an approximate KER distribution leads to a satisfactory overall agreement with the experiment. Deviations observed in the intermediate energy part at the most forward (and possibly the most backward) angles may stem from the different approximations of the model in which a simple Gaussian KER distribution is introduced and in which the vibrational motion of the $H_2O$ molecular target is neglected within the initial conditions. In the high-energy part, the introduced KER increases the width of the simulated recoil peaks, so that the calculated $H^-$ and $H^+$ recoil peaks reproduce fairly well the experimental ones. We note that a KER of a few eV can indeed significantly broaden the H recoil peaks, even if these peaks appear at energies of several hundreds of eV [46]. Moreover, at low emission energy the introduced KER strongly enhances the simulated cross sections, so that the calculated curves agree satisfactorily with the data points obtained from TOF measurements. According to the present simulations, without kinetic-energy release due to electronic excitation or ionization, both total cross sections for $H^+$ and $H^-$ emission would be about three times smaller. We therefore conclude that – via subsequent kinetic-energy release – electronic excitation and ionization processes play a crucial role in both $H^+$ and $H^-$ formation.

### B. Thermodynamic model

In order to explain in more detail the processes behind the observed fragment spectra, we used a recently developed model for molecular collisions [39], which is able to provide absolute cross sections for emission of fragment ions including anions.

Briefly, the model attributes thermal probabilities for the different transient excited states of the collisional quasimolecule. The energy levels of the quasimolecule are obtained from *ab initio* calculations when available or otherwise estimated within the framework of a screened charge potential approximation, and the motion of the atomic centers is determined classically. More information about the model can be found in [39].

In the present calculations slight modifications were implemented. The vibrational motion of the $H_2O$ molecular target was taken into account within the initial conditions of the simulation. This has a significant effect on the energy distribution of the fragments of the target. The energy of the fragments is primarily determined by the internuclear distances between the cores of the molecule, which are significantly influenced by the vibrations. Therefore, the initial positions and velocities of atomic centers of the $H_2O$ target molecule were chosen considering the three vibrational modes of the molecule with random phase and energy equal to the corresponding quantum mechanical zero-point energy. Another change in the model was performed in the applied screened charge potential approximation for the electronic energy levels of the quasi molecule. For excited states, where *ab initio* potentials were not available, the screening radii of the atoms or atomic ions were obtained by Slater's rules [47]. The potential curves for the first few levels of the diatomic subsystems, however, were taken from the literature (see references in [39]).

The obtained H$^-$ and H$^+$ fragment spectra at different observation angles are plotted in Fig. 7 in comparison with the experimental results. The shape of the obtained curves shows a high degree of similarity on the low-energy side. On an absolute scale, theory and experiment agree within a factor of 2 for H$^+$ emission, but the calculated H$^-$ cross sections are about one order of magnitude lower than the experimental ones.

For smaller systems including two or three atoms, better agreement was found earlier for anion production [39]. Here, besides larger complexity of the system, the reason for the discrepancy is probably due to the inaccuracies in the applied screened charge potential approximation for those pair potential energy curves which involve the H$^-$ anion. More precise calculations would require better approximations of the potential curves. Unfortunately, the resulting longer computational times currently exceed our computational limits.

Note that the statistics in the binary peaks is low due to the limited number of trajectories (96 000). Thus, binary processes will not be further discussed within this model. Moreover, the high-energy part of the spectra in Fig. 7 generally suffers from poor statistics, which results in apparent dips or peaks in some cases.

For H$^+$ formation, it seems that the model describes well the dissociation probability of the target and the KER obtained by the fragments, which have to be introduced by hand in the four-body scattering calculation. Moreover, the model enables one to distinguish between the different output channels. The results of the model show that the overwhelming majority of the produced H$^+$ ions stems from a transfer-ionization process in which an electron is captured by the projectile and an electron is removed from the target to the continuum at the same time. The latter ionization process may be viewed as an excitation to the continuum. Moreover, the ionized molecular target may be left in an electronic excited state. These "excitations" contribute to the KERs of the ejected fragments. The present calculations show that transfer ionization has a high probability of 0.7 within 1 a.u. of the impact parameter and it declines fast to zero as the impact parameter further increases. Pure single-capture and single-ionization processes are presently found to give negligible contribution to the fragment-ion spectra. This is reasonable since those processes leave behind a singly ionized target which does not dissociate when created in its ground state. Multiple ionization or capture processes are found to be rare events in the calculations.

The results of this statistical-type model exhibit a high level of similarity between the shapes of the anion and cation spectra, as observed experimentally. It confirms our conjecture that the charge state of the emitted fragment ions is governed by simple statistical laws [36-38]. In the model, this statistical aspect results from thermal population of electronic energy levels of the collisional complex. The practical upper limit of the excitation energy that needs to be considered in the calculations is determined by the characteristic temperature of the thermal distribution. Within the simulations, this temperature nearly reaches $9.3 \times 10^4$ K (80 eV) at maximum in close-approach collisions. This may give a hint for other possible *ab initio* calculations, too, that the sufficient set of electronic energy levels describing collision-induced ion formation may extend to 80 eV above the ground state in case of H$_2$O fragmentation.

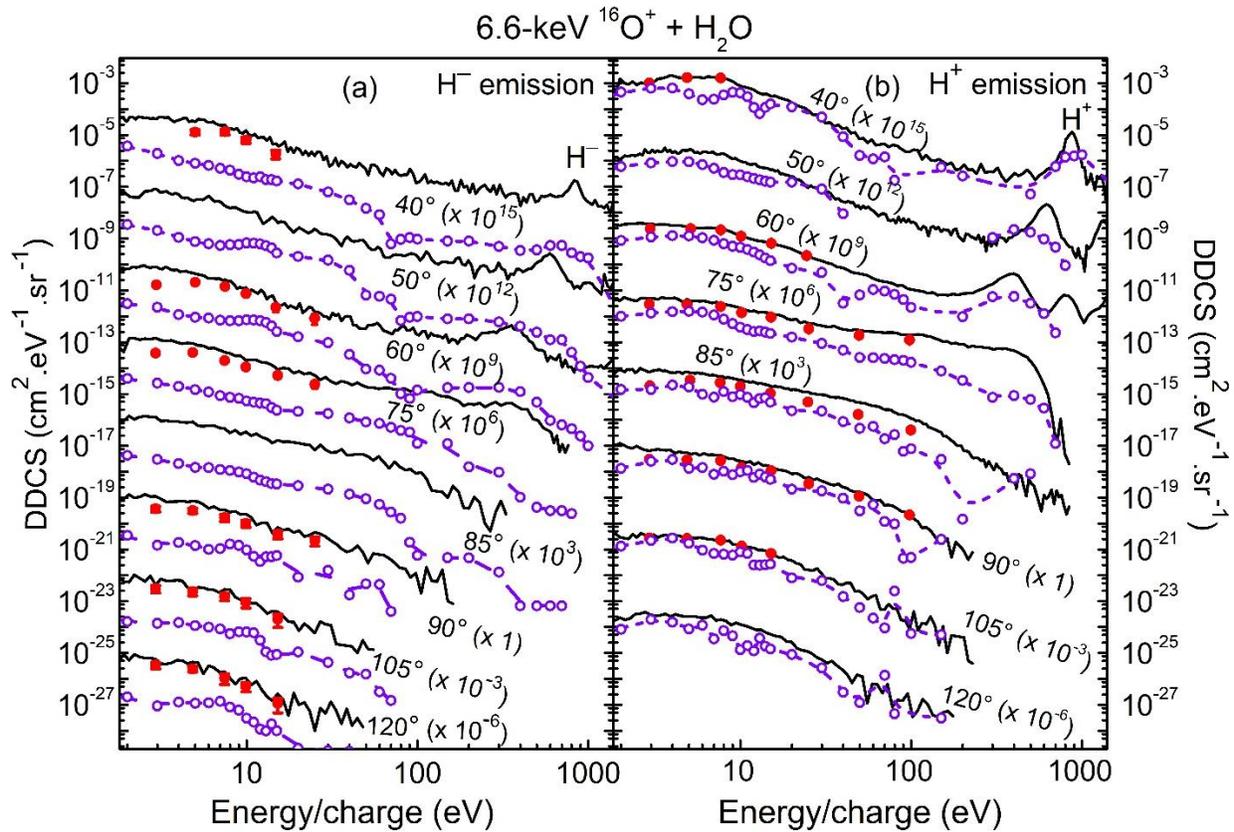

**Figure 7.** (Color online) Comparison between experimental DDCS and calculations by the thermodynamic model for hydrogen fragment emission in 6.6-keV $^{16}O^+$ + $H_2O$ collisions at the indicated observation angles. **Open circles:** Calculated energy distribution of the $H^-$ (a) and $H^+$ (b) fragments obtained within the framework of the thermodynamic model; dashed lines are to guide the eye. **Full curve:** Experimental DDCS for anion (a) and cation (b) emission. **Red, full circles:** Experimental $H^-$ (a) and $H^+$ (b) contributions from TOF measurements. Each spectrum is multiplied by the indicated factor.

## V. MANY-BODY PROCESSES VERSUS COULOMB EXPLOSION IN DIFFERENT COLLISION SYSTEMS

In this section we focus on the mechanisms underlying the formation of low-energy cation fragments. Prior to this work, fragmentation of water by high-energy singly charged projectiles and low-energy highly charged ions was compared and the DDCS fragment spectra were found to be similarly structured [29]. The observed peak structures were explained by Coulomb explosion of the multiply charged target molecule left after the collision, even though the ionization mechanisms were different in the two cases.

Peaks may appear in energy spectra when the electron removal is sudden (transitions of Franck-Condon type during which the target nuclei can be considered as fixed) and when the motion of the ejected fragments is not significantly affected by their interaction with the projectile. The higher the charge state of the ionized $H_2O$ target, the higher the average kinetic energy of the ejected $H^+$ fragments. Multiple ionization of the target can result from multiple electron capture processes in collisions involving slow, highly charged ions. Multiple electron removal can also be induced by ionization and excitation processes in fast collisions, at velocities larger than 1 a.u.

The presently obtained energy spectra of cations show a striking difference from those obtained in collisions involving projectiles of higher charge or higher energy [25-29], as shown in Fig. 8. Here, comparison is made with 650-keV $N^+$ and 70-keV $O^{7+}$ projectiles in the fragment energy range for soft processes at observation angles close to 45°. In the present case with 6.6-keV $O^+$ projectiles, no peak structures are visible in the fragment spectra. This is the signature of many-body processes, which smear out the structures. Namely, the projectile ion as a third body may have significant influence on the Coulomb exploding pair of ionized target fragments. This is supported by the thermodynamic model as it shows that fragmentation is mostly due to transfer ionization at relatively small impact parameter collisions below 1 a.u. On the contrary, multiple electron capture by highly charged ions occuring in large impact parameter collisions and multiple ionization by fast dressed projectiles imply a brief interaction with the target, so that the projectile has little influence on the Coulomb explosion of the target in these cases. The latter is true even if, according to classical trajectory model calculations (CTMC), multiple ionization induced by fast, singly charged projectiles results from close collision events at small impact parameters [29].

We note that the impact parameter of 1 a.u. is still significantly larger than the typical one for binary processes. (The peaks due to direct binary processes appear at significantly higher energies than those shown in Fig. 8.) However, in the impact parameter range around 1 a.u., the direct momentum transfer to the target nuclei is comparable to that gained from Coulomb explosion. Hence, though not as strongly as in binary collisions, direct collisional momentum transfer can significantly influence the spectra presented in Fig. 8.

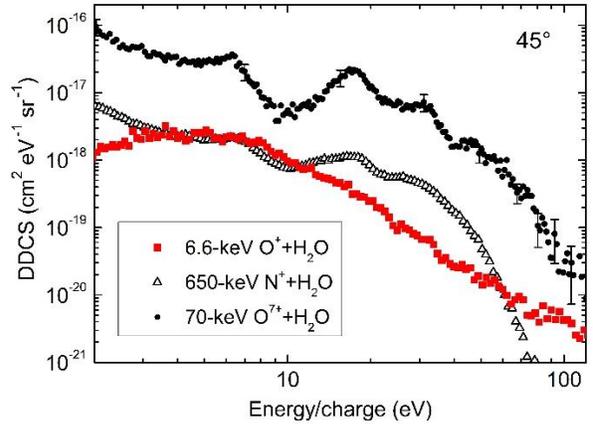

**Figure 8.** Energy distribution of the $H^+$ cations ejected from $H_2O$ by ion impact at 45° (50° for $O^+$ projectile). **Red squares:** for 6.6-keV $O^+$ +$H_2O$ collision, present work. **Open triangles:** for 650-keV $N^+$ +$H_2O$ collision from [29]. **Circles:** for 70-keV $O^{7+}$ +$H_2O$ collision from [29].

It is surprising that collisions with low-energy singly charged ions lead to significant $H^+$ emission above 60 eV because such high-energy $H^+$ ions are associated with the removal of five or six electrons from the target $H_2O$ molecule in the case of high-energy or multiply charged ions [27,29]. However, as suggested by the thermodynamic model, slow, singly charged ions are unlikely to eject so many electrons from the target. Instead, the high fragment energies may stem from the excitation of the target or by the influence of the projectile on the fragmentation rather than from a pure Coulomb explosion (see above). We also note that anion emission cannot result from a pure Coulomb explosion of a positively charged complex. This interpretation in terms of target excitation and momentum transfer from the projectile is also supported by the structural similarity of both the measured and the calculated cation and anion spectra (see Figs. 2, 6, and 7).

## VI. SUMMARY AND CONCLUDING REMARKS

We have investigated the formation of anion and cation fragments in 6.6-keV $^{16}O^+$ + $H_2O$ collisions, both experimentally and theoretically. The experimental setup allowed the measurement of absolute cross sections for the emission of all positively and negatively charged ion fragments differential in energy and observation angle. The most striking finding is that the entire kinetic-energy and angular distributions of the cation fragments are very similar to those of the ejected anions. The main component of the energy distribution of the fragments is broad and slowly decreasing with energy. These fragment ions originate from soft many-body processes. At forward angles ($< 90°$), the pronounced peaks observed at higher energies are due to recoil fragments formed in hard binary collisions occurring at small impact parameters.

The fact that the double-differential cross sections for $H^+$ and $H^-$ formation are nearly proportional at all angles and over the entire emission energy range shows that the relative populations of the different charge states of the hydrogen fragments do not depend significantly on the emission angle, the impact parameter, or the momentum transferred between the collision partners. This finding suggests that the charge-state distribution of the hydrogen fragments is akin to a statistical distribution, independently of whether these fragments are formed via binary or many-body processes.

To interpret the experimental findings, we performed model calculations with a full description of the core-core interactions, including the kinetic energy released in electronic excitation and ionization processes. Simulations within a simplistic four-body scattering model suggest that electronic excitation and/or ionization processes play a crucial role in both $H^+$ and $H^-$ formation.

For a deeper insight into the fragmentation mechanisms and their related electronic processes, we performed further calculations within the statistical thermodynamic model. These calculations reproduce the characteristic features of the doubly differential spectra. Moreover, the model shows that the overwhelming majority of the produced $H^+$ ions stems from transfer-ionization processes occurring at impact parameters smaller than 1 a.u. According to the model, transfer ionization mainly occurs in soft many-body processes in which $H_2O$ fragmentation by slow $O^+$ impact is not exclusively governed by the Coulombic intramolecular repulsion following electron removal, but also by short-distance interactions between the projectile and the atomic cores of the target. This explains why the energy spectra of the fragments exhibit a smooth distribution contrary to the case of fragmentation by high-charge-state or high-energy projectile ions.

Lastly, we note that it is important to treat anion emission as a standard channel in molecular collisions, even if its yield is relatively small. This is relevant in many cases since slow anions are generally highly reactive species that may chemically affect the media in which they are formed.

## ACKNOWLEDGMENTS

We are grateful to C. Feierstein, S. Guillous, A. Leprévost, L. Maunoury, F. Noury, J.-M. Ramillon, P. Rousseau and V. Toivanen for technical assistance. This work was supported by the Hungarian Scientific Research Fund (OTKA Grant No. K128621), the French–Hungarian Cooperation Program PHC Balaton (Grant No. 38620NH), the National Research, Development and Innovation Office (NRDIO) from the Hungarian NRDI Fund (Grant No. TéT_16-1-2016-2026), and by the International Scientific Cooperation Project PICS-CNRS (Grant No. 7739 & PROJEKT 2016-2). J.A.T. was supported by a grant from the Basse-Normandie region (Convention No. 11P01476) and from the European Regional Development Fund (Grant No. 32594).

[1] M. E. Rudd, T.V. Goffe, R. D. DuBois, and L. H. Toburen, Cross sections for ionization of water vapor by 7–4000-keV protons, Phys. Rev. A **31**, 492 (1985).

[2] U. Werner, K. Beckord, J. Becker, and H. O. Lutz, 3D Imaging of the Collision-Induced Coulomb Fragmentation of Water Molecules, Phys. Rev. Lett. **74**, 1962 (1995).

[3] U. Werner, K. Beckord, J. Becker, H. O. Folkerts and H. O. Lutz, Ion-impact-induced fragmentation of water molecules, Nucl. Instr. and Meth. B **98**, 385 (1995).

[4] F. Gobet, B. Farizon, M. Farizon, M. J. Gaillard, M. Carré, M. Lezius, P. Scheier, and T. D. Märk, Total, Partial, and Electron-Capture Cross Sections for Ionization of Water Vapor by 20–150 keV Protons, Phys. Rev. Lett. **86**, 3751 (2001).

[5] F. Gobet, S. Eden, B. Coupier, J. Tabet, B. Farizon, M. Farizon, M. J. Gaillard, M. Carré, S. Ouaskit, T. D. Märk, and P. Scheier, Ionization of water by (20–150)-keV protons: Separation of direct-ionization and electron-capture processes, Phys. Rev. A **70**, 062716 (2004).

[6] H. Luna and E. C. Montenegro, Fragmentation of Water by Heavy Ions, Phys. Rev. Lett. **94**, 043201 (2005).

[7] H. Luna, A. L. F. de Barros, J. A. Wyer, S. W. J. Scully, J. Lecointre, P. M. Y. Garcia, G. M. Sigaud, A. C. F. Santos, V. Senthil, M. B. Shah, C. J. Latimer, and E. C. Montenegro, Water-molecule dissociation by proton and hydrogen impact, Phys. Rev. A **75**, 042711 (2007).

[8] S. Legendre, E. Giglio, M. Tarisien, A. Cassimi, B. Gervais, and L. Adoui, Isotopic effects in water dication fragmentation, J. Phys. B **38**, L233 (2005).

[9] A. M. Sayler, M. Leonard, K. D. Carnes, R. Cabrera-Trujillo, B. D. Esry and I. Ben-Itzhak, Preference for breaking the O–H bond over the O–D bond following HDO ionization by fast ions, J. Phys. B **39**, 1701 (2006).

[10] F. Alvarado, R. Hoekstra, and T. Schlathölter, Dissociation of water molecules upon keV $H^+$- and $He^{q+}$-induced ionization, J. Phys. B **38**, 4085 (2005).

[11] Z. D. Pešić, J.-Y. Chesnel, R. Hellhammer, B. Sulik, and N. Stolterfoht, Fragmentation of $H_2O$ molecules following the interaction with slow, highly charged Ne ions, J. Phys. B **37**, 1405 (2004).

[12] R. Cabrera-Trujillo, E. Deumens, Y. Öhrn, O. Quinet, J. R. Sabin, and N. Stolterfoht, Water-molecule fragmentation induced by charge exchange in slow collisions with $He^+$ and $He^{2+}$ ions in the keV-energy region, Phys. Rev. A **75**, 052702 (2007).

[13] Z. Juhász, B. Sulik, F. Frémont, A. Hajaji, and J.-Y. Chesnel, Anisotropic ion emission in the fragmentation of small molecules by highly charged ion impact, Nucl. Instr. and Meth. B **267**, 326 (2009).

[14] Z. P. Wang, P. M. Dinh, P.-G. Reinhard, E. Suraud, G. Bruny, C. Montano, S. Feil, S. Eden, H. Abdoul-Carime, B. Farizon, M. Farizon, S. Ouaskit, and T. D. Märk, Microscopic studies of atom–water collisions, Int. J. of Mass Spectr. **285**, 143 (2009).

[15] J. Rajput and C. P. Safvan, Orientation and alignment effects in ion-induced fragmentation of water: A triple coincidence study, J. Chem. Phys. **141**, 164313 (2014).

[16] S. Martin, L. Chen, R. Brédy, J. Bernard, and A. Cassimi, Fragmentation of doubly charged HDO, $H_2O$, and $D_2O$ molecules induced by proton and monocharged fluorine beam impact at 3 keV, J. Chem. Phys. **142**, 094306 (2015).

[17] P. M. Y. Garcia, G. M. Sigaud, H. Luna, A. C. F. Santos, E. C. Montenegro, and M. B. Shah, Water-molecule dissociation by impact of $He^+$ ions, Phys. Rev. A **77**, 052708 (2008).

[18] A. L. F. de Barros, J. Lecointre, H. Luna, M. B. Shah, and E. C. Montenegro, Energy distributions of $H^+$ fragments ejected by fast proton and electron projectiles in collision with $H_2O$ molecules, Phys. Rev. A **80**, 012716 (2009).

[19] C. Illescas, L. F. Errea, L. Méndez, B. Pons, I. Rabadán, and A. Riera, Classical treatment of ion-$H_2O$ collisions with a three-center model potential, Phys. Rev. A **83**, 052704 (2011).

[20] M. Murakami, T. Kirchner, M. Horbatsch, and H. J. Lüdde, Single and multiple electron removal processes in proton–water-molecule collisions, Phys. Rev. A **85**, 052704 (2012).

[21] M. Murakami, T. Kirchner, M. Horbatsch, and H. J. Lüdde, Fragmentation of water molecules by proton impact: The role of multiple electron processes, Phys. Rev. A **85**, 052713 (2012).

[22] M. Murakami, T. Kirchner, M. Horbatsch, and H. J. Lüdde, Quantum-mechanical calculation of multiple electron removal and fragmentation cross sections in $He^+$-$H_2O$ collisions, Phys. Rev. A **86**, 022719 (2012).

[23] L. Gulyás, S. Egri, H. Ghavaminia, and A. Igarashi, Single and multiple electron removal and fragmentation in collisions of protons with water molecules, Phys. Rev. A **93**, 032704 (2016).

[24] W. Wolff, H. Luna, R. Schuch, N. D. Cariatore, S. Otranto, F. Turco, D. Fregenal, G. Bernardi, and S. Suárez, Water fragmentation by bare and dressed light ions with MeV energies: Fragment-ion-energy and time-of-flight distributions, Phys. Rev. A **94**, 022712 (2016).

[25] B. Seredyuk, R. W. McCullough, H. Tawara, H. B. Gilbody, D. Bodewits, R. Hoekstra, A. G. G. M. Tielens, P. Sobocinski, D. Pesic, R. Hellhammer, B. Sulik, N. Stolterfoht, O. Abu-Haija, and E. Y. Kamber, Charge exchange and dissociative processes in collisions of slow $He^{2+}$ ions with $H_2O$ molecules, Phys. Rev. A **71**, 022705 (2005).

[26] P. Sobocinski, Z. D. Pešić, N. Stolterfoht, B. Sulik, S. Legendre, and J.-Y. Chesnel, Fragmentation of water molecules in slow $He^{2+}$ + $H_2O$ collisions, J. Phys. B **38**, 2495 (2005) and references therein.

[27] P. Sobocinski, Z.D. Pešić, R. Hellhammer, D. Klein, B. Sulik, J.-Y. Chesnel, and N. Stolterfoht, Anisotropic proton emission after fragmentation of $H_2O$ by multiply charged ions, J. Phys. B **39**, 927 (2006).

[28] S. T. S. Kovács, P. Herczku, Z. Juhász, L. Sarkadi, L. Gulyás, and B. Sulik, Dissociative ionization of the $H_2O$ molecule induced by medium-energy singly charged projectiles, Phys. Rev. A **96**, 032704 (2017).

[29] S. T. S. Kovács, P. Herczku, Z. Juhász, L. Sarkadi, L. Gulyás, J.-Y. Chesnel, L. Ábrók, F. Frémont, A. Hajaji, and B. Sulik, Dissociative ionization of the $H_2O$ molecule bombarded by swift singly charged and slow highly charged ions, X-Ray Spectrometry (2019).

[30] O. Yenen, D. H. Jaecks and L. M. Wiese, Energy distribution of $H^-$ from the collision-induced three-particle breakup of $H_3^+$, Phys. Rev. A **39**, 1767 (1989).

[31] G. Jalbert, L. F. S. Coelho and N. V. de Castro Faria, $H^-$ formation from collisional destruction of fast $H_3^+$ ions in noble gases, Phys. Rev. A **46**, 3840 (1992).

[32] M. Tybislawski, M. Bends, R. J. Berger, M. Hettlich, R. Lork and W. Neuwirth, Production of negatively charged fragments in collisions of swift positive hydrogen ions with molecules, Z. Phys. D **28**, 49 (1993).

[33] F. B. Alarcón, B. E. Fuentes, and H. Martínez, Absolute cross section measurements for dissociative capture of $H_2^+$ in Ar, Int. J. Mass Spectrom. **248**, 21 (2006).

[34] B. Li, X. Ma, X. L. Zhu, S. F. Zhang, H. P. Liu, W. T. Feng, D. B. Qian, D. C. Zhang, L. Chen, R. Brédy, G. Montagne, J. Bernard and S. Martin, High negative ion production yield in 30 keV $F^{2+}$ + adenine ($C_5H_5N_5$) collisions, J. Phys. B **42**, 075204 (2009).

[35] E. J. Angelin and R. Hippler, Superoxide-anion formation in collisions of positively charged argon ios with oxygen molecules, Phys. Rev. A **87**, 052704 (2013).

[36] Z. Juhász, B. Sulik, J. Rangama, E. Bene, B. Sorgunlu-Frankland, F. Frémont, and J.-Y. Chesnel, Formation of negative hydrogen ions in 7-keV $OH^+$ + Ar and $OH^+$ + acetone collisions: A general process for H-bearing molecular species, Phys. Rev. A **87**, 032718 (2013).

[37] E. Lattouf, Z. Juhász, J.-Y. Chesnel, S. T. S. Kovács, E. Bene, P. Herczku, B. A. Huber, A Méry, J.-C. Poully, J. Rangama, and B. Sulik, Formation of anions and cations via a binary-encounter process in $OH^+$ + Ar collisions: The role of dissociative excitation and statistical aspects, Phys. Rev. A **89**, 062721 (2014).

[38] J.-Y. Chesnel, Z. Juhász, E. Lattouf, J. A. Tanis, B. A. Huber, E. Bene, S. T. S. Kovács, P. Herczku, A. Méry, J.-C. Poully, J. Rangama, and B. Sulik, Anion emission from water molecules colliding with positive ions: Identification of binary and many-body processes, Phys. Rev. A **91**, 060701(R) (2015).

[39] Z. Juhász, Thermodynamic model for electron emission and negative- and positive-ion formation in keV molecular collisions, Phys. Rev. A 94, 022707 (2016).

[40] F. Frémont, A. Hajaji, and J.-Y. Chesnel, K-shell and total ionization cross sections following electron-molecule collisions: An empirical scaling law, Phys. Rev. A **74**, 052707 (2006).

[41] F. Frémont, A. Hajaji, R. O. Barrachina, and J.-Y. Chesnel, Interférences de type Young avec une source à un seul électron, C. R. Phys. 9, 469 (2008).

[42] N. Stolterfoht, J.-Y. Chesnel, M. Grether, B. Skogvall, F. Frémont, D. Lecler, D. Hennecart, X. Husson, J.-P. Grandin, B. Sulik, L. Gulyás, and J. A. Tanis, Two- and three-body effects in single ionization of Li by 95 MeV/u $Ar^{18+}$ ions: Analogies with photoionization, Phys. Rev. Lett. **80**, 4649 (1998).

[43] P. Rawat, V. S. Prabhudesai, G. Aravind, M. A. Rahman and E. Krishnakumar, Absolute cross sections for dissociative electron attachment to $H_2O$ and $D_2O$, J. Phys. B **40**, 4625 (2007).

[44] MOLPRO package of ab initio programs designed by H.-J.Werner and P. J. Knowles, version 2010.1. See http://www.molpro.net/.

[45] D. Mathur, Multiply charged molecules, Physics Reports **225**, 193 (1993).

[46] As an example, for a KER of 5 eV, the velocity $v_{KER}$ (~$1.4 \times 10^{-2}$ a.u.) is about 10% of the velocity $v$ of an H fragment with recoil energy of 500 eV. In this case, adding the component $\vec{v}_{KER}$ to the velocity $\vec{v}$ leads to a variation in recoil energy that can reach ~ 20%, thus leading to a significant broadening of any H recoil peak at 500 eV.

[47] J. C. Slater, Atomic Shielding Constants, Phys. Rev. **36**, 57 (1930).